# Automatically Determining a Network Reconnaissance Scope Using Passive Scanning Techniques

Stefan Marksteiner, Bernhard Jandl-Scherf, and Harald Lernbeiß

JOANNEUM RESEARCH, Graz, Austria,
`stefan.marksteiner@avl.com`

**Abstract.** The starting point of securing a network is having a concise overview of it. As networks are becoming more and more complex both in general and with the introduction of IoT technology and their topological peculiarities in particular, this is increasingly difficult to achieve. Especially in cyber-physical environments, such as smart factories, gaining a reliable picture of the network can be, due to intertwining of a vast amount of devices and different protocols, a tedious task. Nevertheless, this work is necessary to conduct security audits, compare documentation with actual conditions or find vulnerabilities using an attacker's view, for all of which a reliable topology overview is pivotal. For security auditors, however, there might not much information, such as asset management access, be available beforehand, which is why this paper assumes network to audit as a complete black box. The goal is therefore to set security auditors in a condition of, without having any a priori knowledge at all, automatically gaining a topology oversight. This paper describes, in the context of a bigger system that uses active scanning to determine the network topology, an approach to automate the first steps of this procedure: passively scanning the network and determining the network's scope, as well as gaining a valid address to perform the active scanning. This allows for bootstrapping an automatic network discovery process without prior knowledge.

## 1 Introduction

As today's networks' complexity, especially with the introduction of *Internet of Things (IoT)*-related technologies and, protocols and architectures with its snares [3, 12] , is becoming higher, the possibility of structural security vulnerabilities also rises. In order to discover these vulnerabilities, it is beneficial not only to use documentation and verify known countermeasures, which would only lead to already paved roads, the usage of black box techniques, resembling an attack could be beneficial [17]. To do so, a network's structure must be uncovered first to reveal weak spots [14]. With said network complexity, however, this can be a tedious task. It is therefore crucial to automate as much of this process as possible in order to yield suitable results with workable effort. There is already work on automated tool-chain based scanning that is dedicated to achieve this task [13] . Based on this work, this paper shows an approach to automatically determining a network's scope to prepare the ground for automated network scanning without any given information. This way, a (possibly external) auditor might be able to black box-map an unfamiliar network by hitting a single button, not using any a priori knowledge.

## 2   Related Work

More than three decades ago it became apparent that the original two-level hierarchy of the Internet was no longer adequate for technical as well as organizational reasons [15]. In 1985, the Internet Engineering Task Force (IETF) specified the rules on how to integrate the new third (*subnet*) layer [16]. Since then, a host that gets attached to a subnet of the Internet needs to know two additional parameters besides its Internet address to be able to communicate with hosts outside the own subnet: the subnet mask and the address of a gateway that connects the subnet with other parts of the Internet. The earlier version [15] already envisioned two major ways for providing the subnet mask. One is *hardwired* (a priori) knowledge (such as reading from persistent configuration), the other is an extension to the ICMP protocol that allows for dynamic determination. This extension was then specified in the form of the *Address Mask Request/Response* message pair in [16]. Together with the *Information Request/Reply* [19] and the *Router Discovery* ICMP messages [9], the scene could be perfectly set for dynamic discovery of the three required communication interface parameters.

Today, the *Information Request/Reply* and the *Address Mask Request/Response* messages are deprecated [10] and the *Router Discovery* messages were never widely implemented [1]; the DHCP protocol [8] is used instead. In scenarios lacking services for dynamic configuration (like no DHCP server present or the DHCP server not willing to lease addresses to unknown hosts), procedures for automatic configuration [11] must be used. The IETF Zeroconf Working Group addressed automatic IP interface configuration as one of their requirements [23]. The working group defined how to obtain a so called *link-local* IPv4 address [7]. But link-local addresses are not suitable for communication with devices not directly connected to the same physical (or logical) link; thus they cannot be used for network mapping beyond that link.

If obtaining a routable address is difficult, why not just passively listen to network traffic? In [18], it is shown that a wealth of information can be gathered by observing mDNS messages. However, approaches like this require services that announce themselves and cannot reach beyond their vLAN on their own. Several Drafts up to number 15 [1] of the standard to detect network attachment [2] suggested (in an appendix) to listen for the network traffic caused by several protocols to make an *educated guess* as to which network a device has moved to. Although used in a different scenario, the idea of listening passively for packets of relevant protocols points into the right direction.

## 3   Methodology

The implemented network reconnaissance procedure can be divided into three main phases. Within the first phase, passive scanner modules observe the network traffic and gather information about hosts in the network. The second phase consists of an analyzer module processing the results from the passive scanners and determining network ranges that will serve as input for subsequent active scanners, which form the third phase (that is outside this paper's scope). The ultimate scope is to black-box detect a network where the analyst, e.g. an auditor performing a security audit or a penetration-testing consultant, discovers the given network without any a priori knowledge. The point of origin of the scan is therefore some point in the local area network (e.g. a given network port to plug in). Figure 1 shows an overview of this process. In case the network device configuration does not fit to any determined network range, the analyzer module will additionally determine all elements needed for a device reconfiguration before start of the active scanners. The described procedure was implemented as a module for the *Tactical Network Mapping (TNM)* plug-in framework. This framework provides functionality to generate a network graph with only a given target network range as input by augmenting data from a configurable toolchain, consisting of existing scanning tools such as *nmap, amap, Dmitry, zmap* and *snmpwalk* to gather .. from a network, with genuine analytics that try to assess the network's structure [13]. These analytics work mainly through determining the parent (or gateway) of each host using *traceroute* information, operating system guessing information and looking for *usual* addresses (such as *.1* or *.254*). Combination of passive scanning and network range determination presented in this paper enhances this framework to be independent of a target range input.

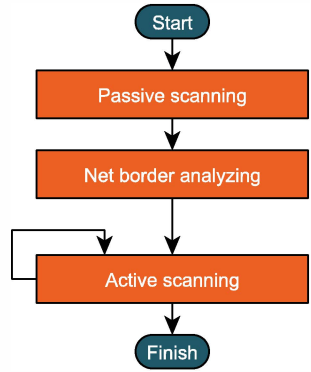

**Fig. 1.** Network reconnaissance procedure

### 3.1 Passive Scanning Phase

The starting point is formed by a couple of parallel executing passive scanners that sniff the network traffic and gather information about communicating hosts. In order to be able to control the termination of the passive scanning two parameters were introduced. The first parameter specifies a duration timeout and the second a threshold of detected hosts. After reaching one of these limits a network border analyzing module will be started. The execution of scan and analyzing modules occurs within the *TNM* framework. Therefore, plug-ins were implemented for some passive scanners that are available on Linux: *Netdiscover*, *P0f* and *CDPSnarf*. It became evident that the control possibilities and delivered host information were not sufficient for the need of net border analyzing. The reason is that *CDPSnarf* is restricted to messages via the *Cisco Discovery Protocol (CDP)* only (restricting the senders to networks devices and omitting hosts), *P0f* is dedicated to analyzing own active outbound and inbound connections (which will not happen with passive scanning only) and *Netdiscover* is restricted to ARP messages only (leaving out using IP packets). An additional reason is that none of these scanners yield information about the *time-to-live (TTL)* of the packets, which can be used to be compared against a list of standard TTL values in operating systems (e.g. [21])[1]. Hence, an own *Hostdiscover* module was implemented based on the *libpcap*-library. The *Hostdiscover* module is able to sniff ARP and IP packages. While the ARP approach is straightforward in the sense that it basically sets a *permit all ARP packets* filter to pcap, the IP method set an IP filter and additionally filters to let only packets of hop counts of 1, 64 or 128 through. These values are configurable but set as default, for most operating systems have according default TTL values [22], so that received non-filtered packets are likely to come from the same segment and, thus, network, which eventually poses the primary scanning target. For ARP, such a filter is not necessary, as the protocol is non-routable anyway [6]. Using these filters (capturing ARP, IP or both), passing network traffic is captured in promiscuous mode. As a result, the number of ARP requests and replies and the number of IP packages are stored together with the host address and can be accessed by subsequent analyzer modules.

### 3.2 Network Range Determination Algorithm

After termination of the passive scanning phase, the network border analyzing module starts with the determination of preliminary network ranges. Three variants for clustering of detected host addresses are implemented, one of which can be chosen by a setting an option within the software implementation. The first variant is the clustering by considering a maximum network size (again configurable, with a default of 256 hosts). Two hosts will fall into different clusters if the host address range would need a network size greater than a given value. Additionally, two networks will be separated, if the addresses span over private [20]

---

[1] This way, it can be prevented to assume a remote network (with more than one hop away) as current network.

or the dynamic link-local space [7]. The second variant makes the clustering against a presumed network prefix. The third variant tries to put all hosts into one big network, except the special ranges mentioned above. Each variant yields one or more clusters of network addresses ranging from a lowest one as starting point to a highest one as an end point determined by the lowest and highest observed addresses that fits into a given range. If, for instance, the lowest observed address is 192.168.0.2 and the highest is 192.168.1.17 with a maximum size of 256, the first variant would split the observed range into two clusters

The process step compares the current network device configuration with the determined network ranges. If the configured IP address fits to any of the determined network ranges, the subsequent steps needed for reconfiguration of the network device will be skipped and the process will proceed with the definition of the final ranges.

If the network configuration does not fit (or no IP address is configured at all at the chosen scanner interface), the first step is to choose one preliminary range to assume as own network (which also determines the subnet mask). For this purpose, the all determined ranges are ordered by following criteria:

1. Network type (globally reachable [5], private , or dynamic link-local);
2. The number of detected hosts for the range in descending order;
3. The respective starting address in ascending order.

The first element of the ordered network range list will be assumed as own network.

To be able to configure the network device for active scanning, a valid IP address is needed. The basis for the selection of an address is given by the list of detected host addresses for the previously chosen own network. The first undetected address that is located between the first and last address of detected hosts will be taken. If no address can be found within the host addresses, an address before the first or behind the last host will be taken. The overall procedure is aborted if no free address could be found.

Further, a default gateway address is determined, primarily by analyzing ARP statistics originating from modules of the starting passive scanning phase. As a first approach, the most often found sender address of ARP replies will be used. If this does not succeed, the most often found target address of ARP requests will be selected. If no ARP statistics are available very commonly used candidates for default gateway addresses are used, which is the first or last address of the determined network address range. Figure 2 illustrates the whole process.

After the found configuration is set to the network interface, the selection of final network ranges that will be used as input by active modules follows. This phase should give the possibility to expand or condense the preliminary determined ranges. The currently only available implementation at this stage is the adaption of network ranges based on the subnet mask of fitting device configurations.

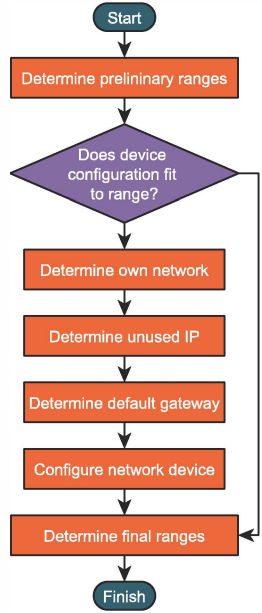

**Fig. 2.** Network border analyzing in detail

### 3.3   Active Scanning Phase

When the final range of the current network is determined, this estimated range or ranges is used as target network for active scanning. Analogously, the found free IP is configured as the host's IP address (see above), in order to obtain a valid one to allow for scanning the network without prior knowledge. The active scanning itself is conducted via an iterative toolchain, containing a variety of scanning tools (such as *nmap*) and analytics modules (such as an algorithm to determine the default gateway of hosts using traceroute, operating system and other information) and yields a topology graph of the target network (i.e. the network, which the host currently resides in). This process, including the gateway determination is described in detail in [13] .

The system was tested in a live, productive network setting. During three test runs, six different networks where found: network $X$ containing the anonymized range $\alpha.\beta.x.0/24$ , network $Y$ containing $\alpha.\beta.y.0/24$ , network $A$ containing $\alpha.\beta.a.0/24$ , network $M$ containing $\alpha.\beta.m.0/24$ , network $N$ containing $\alpha.\beta.n.0/24$ and network $Q$ with the only known address being $\alpha.\beta.q.206$ whereby $X$ and $Y$, as well as $M$ and $N$ are direct neighbors, while $A$ and $Q$ have a range that is disjunct from the other networks. Each test used a toolchain that consisted of the *Hostdiscover* module (see Section 3.1), the network range determination (described in Section 3.2) and an *Nmap* scanner actively scanning the determined range plus a default gateway determination algorithm (see Section 3.3).

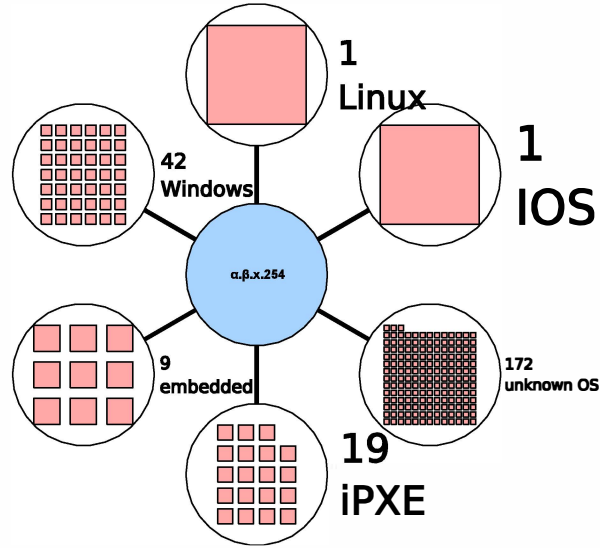

**Fig. 3.** Resulting topology graph from a test run

A first test run, with the scanning host residing in network $X$, used a threshold of 10 found hosts and 5 minutes time for the passive phase. After detection of 10 hosts, which resided in the networks $X$ and $A$, the network range determination correctly identified the ranges for $X$ and $A$ that formed the target for the active phase and was able to discover 244 hosts by determining the network and running an active scan on it (using *Nmap*'s traceroute und OS dection options), 73 from network $X$ and 172 from network $A$. Furthermore, this method automatically yields, next to the network structure, portscanning and operating system detection results for further analysis (see Figure 3 for an anonymized graph of the scan result produced by the *TNM* tool).

A second test with a threshold of 500 hosts and 5 minutes of time, with the scanning host inside network $Y$ yielded in 93 passively detected hosts of which 1 is from network $X$, 51 are from network $Y$ and 41 are from network $A$. As with the first test run, the network ranges where determined correctly using the scanning host's own network mask (see Section 3.2), for its IP address fits in one of the network ranges. Knowing its own home network (Y), it is evident that $X$ and $Y$ are indeed two separate /24 networks and not a single /23 range. The active scanning over the three networks $X$, $Y$ and $A$ (for the sake of time without using *Nmap*'s operating system detection) yielded 309 hosts; 75 from $X$, 63 from $Y$ and 171 from $A$.

A third run with the same setting as the second, found 105 hosts in the passive phase; 50 from network $Y$, 54 from network $A$ and 1 from network $M$. The network range determination correctly identified the networks $Y$ and $A$ but assumed a combined network $M + N$. The separate nature latter, however, became evident in the active scanning. The active phase yielded a total of $253^2$ hosts: 63 from $Y$, 171 from $A$, 13 from $M$, 5 from $N$ and 1 from $Q$. The host from $Q$ is a special case, as it only occurs in traceroute data (residing on the path toward network $N$). Through the different traceroute data, it also became evident that networks $M$ and $N$ where in fact separate networks (the former directly adjacent to $Y$, the latter having an additional hop residing in $Q$). Figure 4 depicts the discovered topology with a comparison graph out of the *TNM* tool that highlights the hosts additional hosts found in the active phase in green, while the uncolored (white for nodes and black for edges) ones where already found during the passive phase.

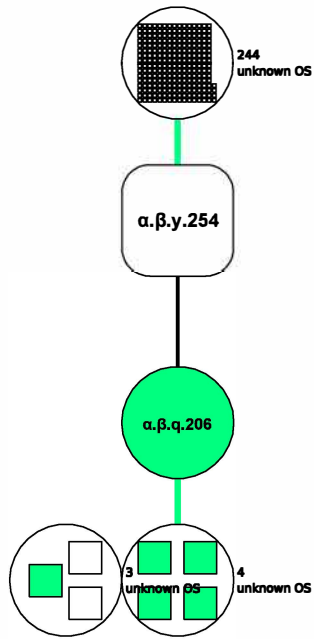

**Fig. 4.** Comparison graph from another test run, highlighting the hosts found in active phase (green)

## 4   Results

The three test runs showed, apart from the knowledge gain regarding the network's structure, a significant increase in discovered hosts by the active phase, compared to simple passive sniffing alone (see Table 1).

**Table 1.** Results of the test runs

| Test Run No. | 1 | 2 | 3 |
|---|---|---|---|
| Passive Results | 10 | 93 | 108 |
| Number of Networks | 2 | 3 | 5 |
| Active Results (including passive) | 244 | 309 | 253 |
| Network $X$ | 73 | 75 | - |
| Network $Y$ | - | 63 | 63 |
| Network $A$ | 172 | 171 | 171 |
| Network $M$ | - | - | 13 |
| Network $N$ | - | - | 5 |
| Network $Q$ | - | - | 1 |
| Detection Gain Ratio | 24.4 | 3.32 | 2.38 |

## 5   Conclusion and Discussion

The work outlined in this paper has shown a path toward automating network scanning without any a priori knowledge. It demonstrated how a network range can determined automatically through network sniffing and automated analysis and be used as a target for an active scanning toolchain. The method showed a significant increase in device detection compared to sole passive sniffing methods, as well as the ability to retrieve topology and other information (such as device types, operating systems, etc.) automatically.

This work has, also uncovered some shortcomings in detecting more segmented (sub)networks. This means that the current process is vital in the sense that a preliminary estimation on the basis of Section 3 is necessary, but might the final determination be enhanced by using an algorithm resembling a binary search [24] that compares traceroute information[3] information of portions[4] of the preliminary ranges, once they exist and the scanning hosts has acquired a valid IP address, refining the results' accuracy. This improvement is subject to continuative works. Furthermore, continuative research will elaborate approaches to handle IPv6 networks.

---

[3] For instance with *nmap -sn -Pn [network_portion]*. This ensures traceroutes to be carried out, even when no active host resides in the network to be examined.

[4] A possibility would be to compare the most distant addresses and, by binary splitting the set, keep comparing until the traceroutes are equal to yield actual subnetworks.

### 5.1 Improvements for gateway determination during the active phase

When a router's internal (that is non-outside) interface residing inside the target range is directly addressed, it yields the same hopcount as the external interface. The rationale is that a router must reduce the initally set *time-to-live (TTL)* value of each packet and must discard it if its decreased to zero, except if is destined to the router itself, where it has to act as a host [4]. In this case, the TTL has no hopcount function (except in the rare case of source routing usage) [6]. That means that the internal interface of a network, assuming that, like any other scanned, it is reachable from the outside, will display a hopcount that is one less than the hopcounts of the other hosts of the scanned network, when viewed from the outside. The missing address, representing this subtracted hopcount, will naturally be the one of the outside interface of this very router.

That allows for determining the internal address of the connecting router closest to the target, which is probably the standard gateway.

## Acknowledgement

This work was partly supported by the Austrian Research Promotion Agency (FFG) within the *ICT of the future* grants program, grant nb. 863129 (project *IoT4CPS*), of the Federal Ministry for Transport, Innovation and Technology (BMVIT) and by the Federal Ministry of Defence (BMLV).